Spin relaxation mechanism in Silver nanowires covered with MgO protection layer

Hiroshi Idzuchi[1,2*], Yasuhiro Fukuma [2,3], Le Wang [2, 4], and Yoshichika Otani [1, 2, 4]

[1] *Institute for Solid State Physics, University of Tokyo, Kashiwa 277-8581, Japan*

[2] *Advanced Science Institute, RIKEN, 2-1 Hirosawa, Wako 351-0198, Japan*

[3] *Department of Computer Science and Electronics, Kyushu Institute of Technology, 680-4 Kawazu, Iizuka 820-8502, Japan*

[4] *Department of Material Physics and Chemistry, University of Science and Technology Beijing, Beijing 100083, People's Republic of China*

Corresponding Author

[*] E-mail: idzuchi@issp.u-tokyo.ac.jp

**Abstract**

Spin-flip mechanism in Ag nanowires with MgO surface protection layers has been investigated by means of nonlocal spin valve measurements using Permalloy/Ag lateral spin valves. The spin flip events mediated by surface scattering are effectively suppressed by the MgO capping layer. The spin relaxation process was found to be well described in the framework of Elliott-Yafet mechanism and then the probabilities of spin-filp scattering for phonon or impurity mediated momentum scattering is precisely determined in the nanowires. The temperature dependent spin-lattice relaxation follows the Bloch-Grüneisen theory and falls on to a universal curve for the monovalent metals as in the Monod and Beuneu scaling determined from the conduction electron spin resonance data for bulk.

Spin injection, transport and detection are key ingredients in spintoronics, which have drawn a great deal of attention in recent years due to possible application for magnetic memories as well as fundamental interests concerning the interplay between charge and spin transport.[1,2] Lateral spin valves (LSVs) offer an effective means to study transport properties of a pure spin current, i.e., a diffusive flow of spin angular momentum accompanying no charge currents. A large number of the spin injection experiments have been reported since the pioneering work by Johnson and Silsbee in 1985.[3] More recently non-local spin valve experiments by Jedema *et al.*[4] brought renewed interests in LSVs in response to the timely development in both micro-fabrication technology and emergent interest in the pure spin current. Having a better insight into the spin transport and relaxation mechanism in nano-scaled devices is important to enhance the performance for the spintronic application.

The spin relaxation mechanism in nonmagnetic metals (NM) has originally been discussed by Elliott and Yafet.[5,6] According to their theory, the spin-orbit interaction (SOI) in NM lifts the spin degeneracy of Bloch electrons, and results in two different energy states for up or down spin. The spin relaxation, i.e., the transition between the opposite spin states, can therefore be caused by the spin independent momentum scatterings due to impurities, grain boundaries, surfaces and phonon.[5-7] The earlier experimental works on the spin relaxation mechanism were mainly performed by conduction electron spin resonance (CESR) measurements and the results were *e.g.*, analyzed by Monod and Benue in the material independent manner.[8,9] Fabian and Das Sarma revisited the CESR analyses and showed a quantitative relationship between the spin flip scattering and the phonon mediated change in resistivity including the effect of Fermi surface topology.[10,11] Since the emergent development in the spintronic devices requires understanding the spin relaxation mechanism in nanowires, the mechanism has been intensively studied by means of nonlocal spin injection techniques using LSVs.[12-17] However, a large contribution of the surface scattering hampered a quantitative analysis of the spin relaxation mechanism in the NM nanowires.[14-16] In this work, we have studied precisely the spin relaxation mechanism due to phonon scattering in Ag nanowires. The surface spin scattering is suppressed by MgO protection layer and then the spin-flip probability for phonon or impurity scattering is determined accurately on the basis of Elliott-Yafet (E-Y) mechanism. The temperature dependence of the spin-lattice relaxation for the phonon scattering is well fitted to the Bloch-Grüneisen formula,[18] showing good consistency with that obtained from CESR measurements in the bulk.

LSVs with $Ni_{80}Fe_{20}$ (Py) / Ag junctions were fabricated with great care. First, suspended resist mask patterns consisting of a bilayer resist, 500 nm-thick methyl methacrylate and 50 nm-thick Poly methyl methacrylate, were formed on a $Si/SiO_2$ substrate by means of e-beam lithography. Then shadow evaporation was performed by using the suspended resist mask to get a clean interface: a Py layer was first e-beam deposited at an angle of 45 degrees from substrate normal, followed by deposition of an Ag layer normal to the substrate cooled by liquid nitrogen. Note here that the ferromagnetic and nonmagnetic materials were deposited separately in the interconnected two different evaporation chambers under ultra-high vacuum condition (~$10^{-7}$ Pa) to prevent degradation of $\lambda_{sf}$ due to magnetic impurities entering the nonmagnetic nanowire. Finally, a MgO capping layer was deposited to avoid surface contamination of the device. The schematic diagram of the LSV fabricated is shown in Figure 1a and the cross-sectional transmission electron micrograph (TEM) of the Ag nanowire and the corresponding energy dispersive X-ray mapping of Ag, Mg and O are shown in Figure 1b. The coverage factor of the entire surface of the wire for the MgO protection layer is estimated to be about 85%. The LSV consists of the Ag nanowire and two Py wires which are electrodes for spin injection and detection. The center to center separation $L$ between the injector and the detector was varied from 300 nm to 1500 nm to determine $\lambda_{sf}$. To study spin flip mechanism for various Ag nanowires, three classes of LSVs are fabricated: 50 nm-thick and 150 nm-wide Ag wire with 3 nm-thick MgO capping layer for device #1, 50 nm-thick and 150 nm-wide Ag wire for device #2 (without MgO), and 100 nm-thick and 200 nm-wide Ag wire with 5 nm-thick MgO capping layer for device #3. The devices #1 and #2 are annealed at 400 °C for 40 min in an $N_2$ (97%) and $H_2$ (3%) atmosphere to improve crystallographic quality of the Ag nanowire.[19, 20] The thickness of Py is fixed at 20 nm, and the width is 120 nm, 120 nm and 150 nm for device #1, #2 and #3, respectively. The resistivity of Py was 4.70×10$^{-5}$ Ωcm and 3.46×10$^{-5}$ Ωcm respectively at 300 K and 5 K.

The nonlocal spin injection measurements were performed on the LSVs with clean Py/Ag junctions. Conventional current-bias lock-in technique was used with applied current amplitude of 0.30 mA and frequency of 79 Hz. The magnetic field was applied parallel to the Py wires. The field dependence of the spin signal for device #1 with $L$ = 300 nm is shown in Figure 1c. Clear spin valve signals $\Delta R_S$ were observed to be 2.45 mΩ at 300 K and 8.92 mΩ at 5 K. Figure 1d displays a reasonable decrease in $\Delta R_S$ with an increase of $L$, attributable to the spin relaxation in the Ag nanowire. Here we assume a transparent interface for the Py/Ag junction of our devices, *i.e.* clean interfaces confirmed by TEM analyses and very low interface resistance of the Py/Ag junction below the resolution ability of 1×10$^{-3}$ Ωμm$^2$ of our measurement system. In this case, the analytical expression of $\Delta R_S$ can be obtained as below by solving the one-dimensional spin diffusion equation for the LSV geometry.[21]

$$\Delta R_S = 4 R_{Ag} \frac{\left(P \frac{R_{Py}}{R_{Ag}}\right)^2 e^{-\frac{L}{\lambda_{Ag}}}}{\left(1 + 2\frac{R_{Py}}{R_{Ag}}\right)^2 - e^{-\frac{2L}{\lambda_{Ag}}}}, \quad (1)$$

where $P$ is the spin polarization of ferromagnet, $R_{Ag}$ = $\rho_{Ag}\lambda_{Ag}/t_{Ag}w_{Ag}$ and $R_{Py}$ = $\rho_{Py}\lambda_{Py}/w_{Py}w_{Py}/(1-P^2)$ are the spin resistances for Ag and Py, respectively, where $\rho$ is the resistivity, $t$ is the thickness, and $w$ is the width. The experimental data were fitted by adjusting parameters $P$ and $\lambda_{Ag}$ with setting the value of $\lambda_{Py}$ = 5 nm reported by Dubois et al.[22] We then obtained $P$ = 0.343±0.025 and 0.485±0.015 and $\lambda_{Ag}$ = 316±28 nm and 851±98 nm, respectively, at 300 K and 5 K.

Figure 2a shows the temperature dependence of $\lambda_{Ag}$. For device #2 (without MgO), $\lambda_{Ag}$ shows maximum at low temperature, which is previously reported for both Cu and Ag nanowires in LSVs due to the surface spin scattering.[14-16] However, monotonic decrease in $\lambda_{Ag}$ with temperature is observed for devices #1 and #3 (with MgO). The MgO capping layer could effectively suppress the surface spin-flip event. The spin relaxation time $\tau_{sf} = \lambda_{Ag}^2 / D_{Ag}$ was calculated by using the diffusion constant $D_{Ag}$ deduced from the Einstein's relation $D_{Ag} = \left( e^2 N(\varepsilon_F) \rho_{Ag} \right)^{-1}$, with the density of states $N(\varepsilon_F)$ = 1.55 states/eV/cm$^3$.[23] The temperature dependence of $\tau_{sf}$ in Figure 2b shows that $\tau_{sf}$ stays almost constant at 16.2 ps and 12.6 ps below 30 K for devices #1 and #3, respectively, indicating that the phonon contribution is frozen out. The temperature increase above 30 K promotes phonon mediated scattering, resulting in a gradual decrease in $\tau_{sf}$. According to the E-Y mechanism, the total spin relaxation time is given by

$$\frac{1}{\tau_{sf}} = \frac{1}{\tau_{sf}^{ph}} + \frac{1}{\tau_{sf}^{imp}} = \frac{\varepsilon_{ph}}{\tau_e^{ph}} + \frac{\varepsilon_{imp}}{\tau_e^{imp}}, \qquad (2)$$

where $\tau_{sf}^{ph}$ and $\tau_{sf}^{imp}$ are spin relaxation times, $\tau_e^{ph}$ and $\tau_e^{imp}$ are momentum relaxation times, and $\varepsilon_{ph}$ and $\varepsilon_{imp}$ are probabilities of spin-flip scatterings. The notations "ph" and "imp" mean respectively phonon and impurity (including grain boundaries) mediated scatterings or probabilities. For devices with the MgO capping layer, since the constant value of $\tau_{sf}$ at the low temperatures is considered as $\tau_{sf}^{imp}$, the temperature variation of $\tau_{sf}^{ph}$ can be deduced from Equation (2). However, for device #2 without the capping layer, the temperature dependence of the surface scattering at low temperatures causes difficulty in determining $\tau_{sf}^{imp}$. Therefore, by assuming $\varepsilon_{imp}$ is same for device #1 and #2, $\tau_{sf}^{imp}$ = 14.2 ps is derived from $\tau_e^{imp}$ at 5 K. Experimental data are deviated from Equation (2) below 40 K as depicted in Figure 2b, meaning that the surface scattering is pronounced. Surface spin-flip is dominated by spin-orbit interaction.[25,26] Thus, the surface spin-flip probability, $\varepsilon_{surf}$, is of the order of the magnitude, $(\alpha Z)^4$, where $\alpha = e^2/\hbar c$ and Z is an atomic number.[25,26] In the present study, the MgO capping layer decreases effective Z at the surface, resulting in suppression of the spin-flip scattering.

The temperature dependence of $1/\tau_{sf}^{ph}$ and $\rho_{ph}$ is analyzed on the basis of E-Y mechanism combined with Bloch-Grüneisen (B-G) theory describing the phonon mediated change in resistivity $\rho_{ph}$ in the entire temperature range. With using Drude model, one obtains

$$\begin{cases} 1/\tau_{sf}^{ph} = \varepsilon_{ph} n e^2 \rho_{ph} / m_e & (3a) \\ \rho_{ph} = Kf(T/\Theta_D) = K\left(\frac{T}{\Theta_D}\right)^5 \int_0^{\Theta_D/T} \frac{z^5 dz}{(e^z-1)(1-e^{-z})}, & (3b) \end{cases}$$

where $m_e$ is the electron mass, $n$ is the free electron density (= $5.86 \times 10^{22}$ cm$^{-3}$ for Ag),[27] $K$ is a constant for a given metal and $\Theta_D$ is Debye temperature. Experimentally obtained phonon contribution to the resistivity $\rho_{ph}$ is fitted to Equation 3b with $K$ and $\Theta_D$ as fitting parameters. Then we obtain $K$ and $\Theta_D$ as $5.15 \times 10^{-6}$ Ωcm and 184 K, $4.70 \times 10^{-6}$ Ωcm and 175 K, and $4.23 \times 10^{-6}$ Ωcm and 184 K for devices #1, #2 and #3, respectively. These are in good agreement with reported values.[28] By comparing $\rho$ with $\tau_{sf}$, we obtain $\varepsilon_{ph}$ and $\varepsilon_{imp}$ for the Ag nanowires in substantial agreement with reported values for the bulk: $\varepsilon_{ph}$ and $\varepsilon_{imp}$ are, respectively, $2.61 \times 10^{-3}$ and $4.03 \times 10^{-3}$ for devices #1 and #2, and $1.60 \times 10^{-3}$ and $3.89 \times 10^{-3}$ for device #3; $\varepsilon_{ph} = 2.86 \times 10^{-3}$ and $\varepsilon_{imp} = 2.50 \times 10^{-3}$ for the bulk from CESR study.[29] The temperature dependence of $1/\tau_{sf}^{ph}$ is in good agreement with B-G curve based on CESR study,[24] as shown in Figure 3a. In this way both LSV and CESR results are found consistent with the B-G theory in the entire temperature range. The nonlocal spin injection technique is useful for determining the spin-flip mechanism at high temperatures.

As in Equation (3b), reduced resistivity $\rho/K = f(T/\Theta_D)$ is material independent. Equation (3a) shows that the spin-lattice relaxation is also material independent after proper scaling by considering the spin-orbit interaction. Based on E-Y mechanism, Monod and Benue estimated a magnitude of the effect of the spin-orbit interaction as $\lambda/\Delta E$ where $\lambda$ is the spin-orbit splitting and $\Delta E$ is the separation to the nearest band with the same transformation properties.[6] Then they found the reduced temperature dependence of $1/\tau_{sf}^{ph}/(\lambda/\Delta E)^2$ in CESR data shows material independent B-G curve for noble and monovalent alkali metals. In Equation (3), by substituting representative value of $(\lambda/\Delta E)^2$ into $\varepsilon$, spin flip rate is expressed as, $C/\tau_{sf}^{ph} \sim Bf(T/\Theta_D)$ where $C$ is the material constant $(\Delta E/\lambda)^2/\gamma_e/\rho_{\Theta_D}$, $\gamma_e = g\mu_B/\hbar$ is the gyromagnetic ratio, $g$ is the g-factor and $\mu_B$ is the Bohr magneton and $B$ is constant $\alpha n e^2 m_e/\gamma_e$. Note that $K$ is expressed as $K = \alpha \rho_{\Theta_D}$ where $\alpha$ is constant 4.225.[28, 30] Revised Monod-Beuneu scaling of $C/\tau_{sf}^{ph}$ vs $T/\Theta_D$ is shown in Figure 3b for CESR data of noble and monovalent alkali metals together with the nonlocal spin injection data for the Ag nanowires.[11] The experimental data in the present study fall well on to the universal curve which reflects intrinsic feature of the Ag nanowires.

In summary, we have investigated the spin relaxation mechanism in Ag nanowires by means of nonlocal spin injection method. The spin-flip probability caused by phonon and impurity scatterings is precisely determined because a surface spin scattering of the Ag wire is well suppressed by an MgO capping layer. This may be due to a decrease of effective $Z$ at the surface. The spin relaxation mechanism can be explained in the framework of the conventional Elliott-Yafet model. The temperature dependence of the spin relaxation times due to phonon scattering for the Ag nanowire is well fitted to the the Bloch-Grüneisen formula and the fitting curve is in good agreement with that of Ag foils used in the CESR measurements. The experiment data in the present study fall on to the universal curve for monovalent metals in the revised Monod-Beuneu scaling. These allow us to determine the spin relaxation time for phonon scattering in the monovalent metals at any temperatures and to understand the spin-flip mechanism in the nonmagnetic nanowires.

**Acknowledgments**

This work was partly supported by Grant-in-Aid for Scientific Research (A) (No. 23244071) and Young Scientist (A) (No. 23681032) from the Ministry of Education, Culture, Sports, Science and Technology, Japan.

**Figure captions**

**Figure 1:** (a) A schematic diagram of Lateral spin valve (LSV) in this study. (b) Cross-sectional transmission electron micrograph of Ag nanowire and corresponding energy dispersive X-ray mapping for Ag, Mg and O. (c) Nonlocal spin signal as a function of magnetic field for device #1 with $L$ = 300 nm at $T$ = 300 K and 5 K. (d) Spin valve signal as a function of injector-detector separation at $T$ = 300 K and 5 K. The solid lines are the fitting curves using Equation (1).

**Figure 2:** (a) Temperature dependence of spin diffusion length of Ag nanowires. Solid lines show fitted curves based on the Elliott-Yafet mechanism. (b) Temperature dependence of spin relaxation time of Ag nanowire. Solid lines show fitted curves for the Elliott-Yafet mechanism. (See text for difference between data and fitted curve for device #2) (c) Temperature dependence of resistivity of Ag nanowire. Solid lines show fitted curves for the Bloch-Grüneisen theory.

**Figure 3:** (a) Temperature dependence of spin relaxation rate *via* phonon scattering. Closed circles show experimental data for the Ag nanowires in the nonlocal spin injection measurement of device #1 (with capping MgO and 50 nm-thick), #2 (without capping MgO and 50 nm-thick) and #3 (with capping MgO and 100 nm-thick) and for Ag foil in the conduction electron spin resonance experiment[24]. Line is the theoretical curve of the Bloch-Grüneisen model. (b) Revised Monod-Beuneu plot, $C/\tau_{sf}^{ph}$ vs $T/\Theta_D$, with experimental data for spin injection and conduction electron spin resonance measurements, where $\Theta_D$ is the Debye Temperature and $C$ is the material constant $\left(\Delta E/\lambda\right)^2 \big/ \gamma_e \big/ \rho_{\Theta_D}$ (see the text for the definition.)

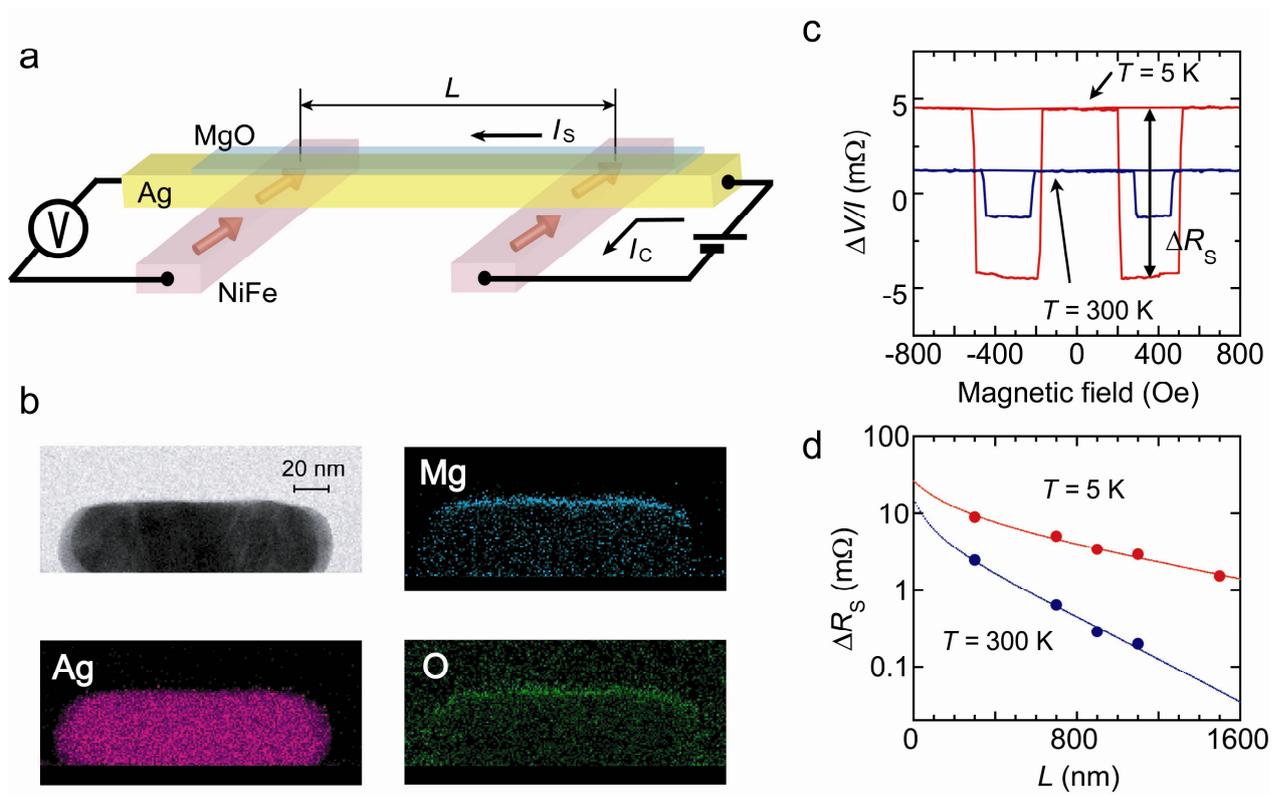

Fig.1 Idzuchi et al.

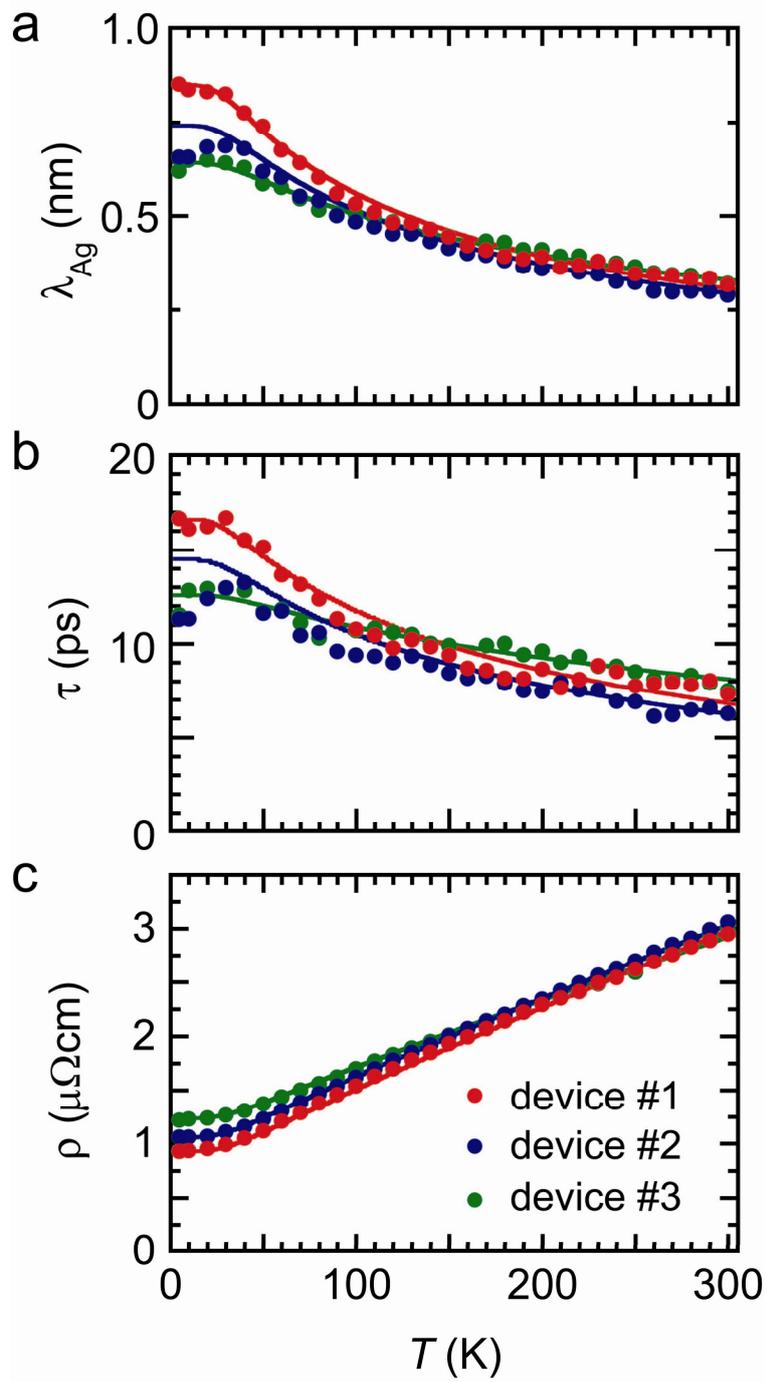

Fig.2　Idzuchi et al.

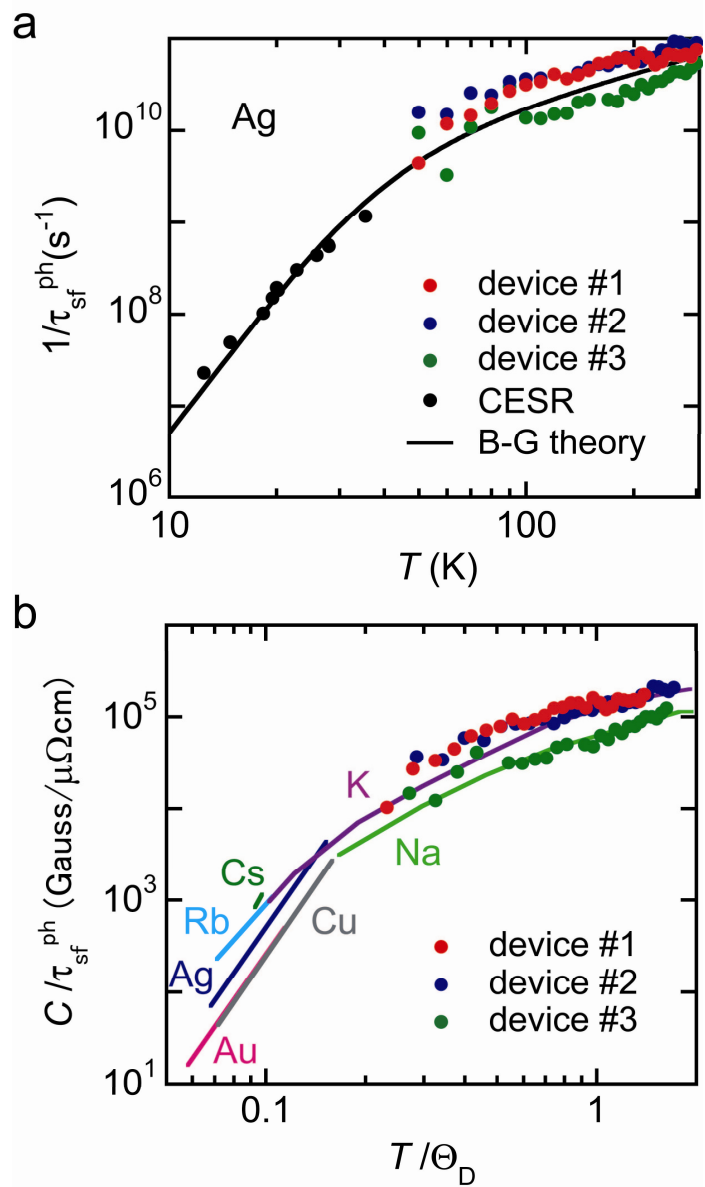

Fig.3 Idzuchi et al.